\documentclass[twocolumn,aps,showpacs,preprintnumbers,amsmath,amssymb,byrevtex,superscriptaddress]{revtex4}
\usepackage{graphicx}
\usepackage{dcolumn}
\usepackage{bm}

\begin{document}
\preprint{Fluorescence Spectrum}
\title{Resonance Fluorescence Squeezing Spectra from Photonic Bandgap Crystals}
\author{Ray-Kuang Lee}
\affiliation{Institute of Electro-Optical Engineering, National Chiao-Tung University, Hsinchu, Taiwan}
\affiliation{National Center for High-Performance Computing, Hsinchu, Taiwan}
\author{Yinchieh Lai}
\email{yclai@mail.nctu.edu.tw}
\affiliation{Institute of Electro-Optical Engineering, National Chiao-Tung University, Hsinchu, Taiwan}
\date{\today}

\begin{abstract}
The fluorescence intensity and quadrature spectra from a two-level atom embedded in a photonic bandgap crystal and resonantly driven by a classical pump light are calculated.
The non-Markovian nature of the problem caused by the non-uniform distribution of the photonic density of states is handled by linearizing the generalized optical Bloch equations with the Liouville operator expansion.
Unlike the case in free space, we find that the bandgap effects will not only modify the fluorescence spectral shape but also cause squeezing in the in-phase quadrature spectra.
\end{abstract}

\maketitle

The study of fluorescence spectra from two-level atoms have been a central topic in quantum optics since the beginning era of quantum mechanics in 1930's.
From the view point of light scattering, both elastic Rayleigh scattering and inelastic Raman scattering processes are 
involved \cite{Loudon83} and thus the fluorescence spectra will have a triplet shape as first calculated by Mollow \cite{Mollow69}.
Theoretical calculations of the fluorescence spectra has been explored \cite{Cohen77,Kimble77} and also verified in experiments\cite{Wu77}.
The squeezing phenomena in the phase-dependent fluorescence spectra of the quadrature field component were first predicted in 1981 by Walls and Zoller \cite{Walls81} and also in 1982 by Mandel \cite{Mandel82}.
It has been theoretically shown that the squeezing can be found in the out-of-phase quadrature component spectra under the condition that $\Omega^2\le \frac{1}{4}\Gamma^2$ \cite{MCollett84, DWalls84}, where $\Omega$ is the Rabi frequency and $\Gamma$ is the atomic decay rate.
Squeezed fluorescence spectra have also been experimentally observed in an experiment using $^{174}$Yb atoms \cite{Lu98}.

In the theoretical studies of fluorescence spectra for the free space case, the Markovian approximation is usually used to describe the statistical properties of the optical noises.
This is a good assumption for the free space case but is not applicable for the case of photonic bandgap crystals \cite{SJohn87,EYablonovitch87}.
This is because in a photonic bandgap crystal, the propagation of light is prohibited within a certain range of wavelengths (the bandgap) due to the lack of available photon states.
Because of the bandgap effects, the distributions of the photonic density of states (DOS) are typically highly non-uniform near the bandedge.
Such a property has prohibited the direct applicability of the Markovian approximation to simplify the derivation for the problem we are going to consider.

In recent years the atom-photon interaction in photonic bandgap materials has been found to exhibit many interesting new phenomena \cite{SJohn90,SJohn94,SYZhu97}.
But almost all of these studies are focused on the transient behavior of the atom-photon interaction and to the best of our knowledge there is still no theoretical treatment on calculating the steady state fluorescence squeezing spectra from photonic bandgap crystals.
The aim of this paper is thus to investigate the properties of the steady state resonance fluorescence emitted by a two-level atom embedded in a photonic bandgap crystal and driven by a classical pumping light.
We treat the photon states in the photonic crystal as the background reservoir and obtain a set of generalized Bloch equations for the atomic operators by eliminating the reservoir field operators.
The non-uniform DOS distributions near the bandedge are modeled by the three-dimensional anisotropic dispersion relation \cite{SZhu00} and the Liouville operator expansion is used to reduce the two-time atomic operator products into equal-time atomic operator products.
In this way the nonlinear generalized Bloch equations are reduced into a set of linear equations with memory function terms caused by the atom-reservoir interaction.
This set of linear equations can then be easily solved in the frequency domain and the resonance fluorescence spectra can be directly obtained from the correlation functions of the atomic operators in the frequency domain.
After performing numerical calculation, we find that the spectral shape of the fluorescence intensity spectra will vary with the wavelength offset between the atomic transition wavelength and the bandedge.
More interestingly, squeezing phenomena are found to present in the in-phase quadrature spectra instead of the out-of-phase quadrature spectra.

To begin the derivation, the Hamiltonian for the system to be considered can be written as:
\begin{eqnarray} 
\label{Hamiltonian} 
H &=& \frac{\hbar}{2}\omega_a\sigma_z+\hbar\sum_k\omega_k a_k^\dag a_k+ \frac{\Omega}{2}\hbar(\sigma_-e^{i\omega_L t}+\sigma_+e^{-i\omega_L t})\nonumber\\ 
&+&\hbar\sum_k(g_k \sigma_+ a_k + g_k^\ast a_k^\dag \sigma_-) 
\end{eqnarray} 
Here the transition frequency of the atom and the frequency of the pumping light are denoted by $\omega_a$ and $\omega_L$ respectively, $a_k^\dag$ and $a_k$ are the creation and annihilation operators of the photon states in the photonic bandgap crystals, $\Omega $ is the Rabi-flopping frequency of the atom under the external pumping light and also represents the relative magnitude of the pumping light, $\sigma_z\equiv(|2\rangle\langle 2|-|1\rangle\langle 1|)$, $\sigma_+\equiv|2\rangle\langle 1|=\sigma_-^\dag$ are the usual Pauli matrices for the two-level atom, and $g_k$ is the atom-field coupling constant. Here we have used the index $k$ to label different photon states and the coupling constant $g_k$ can be expressed as: 
\begin{eqnarray} 
\label{eqgk}
g_k(\hat{\textbf d},\vec{r}_0)\equiv g_k= |d| \omega_a \sqrt{\frac{1}{2\hbar \epsilon_0\omega_k V}}\,\hat{\textbf d}\cdot {\textbf E}^{\ast}_k (\vec{r}_{0}) 
\end{eqnarray} 
Here we use notations $|d|$ for the magnitude of the atomic dipole moment, $\hat{\textbf d}$ for the unit vector along the direction of the dipole moment, $V$ for the volume of the quantization space, and $\epsilon_0$ for the Coulomb constant.

Starting from the Hamiltonian, we treat the photon field operators as the background reservoir and eliminate their corresponding equations to obtain the following set of generalized Bloch equations.
\begin{eqnarray} 
\label{es1}
\dot{\sigma}_-(t) &=& i\frac{\Omega}{2}\sigma_z(t)e^{-i\Delta t}\\\nonumber
&+& \int_{-\infty}^t d\,t' G(t-t')\sigma_z(t)\sigma_-(t') + n_-(t)\\ 
\label{es2} 
\dot{\sigma}_+(t) &=& - i\frac{\Omega}{2}\sigma_z(t)e^{i\Delta t}\\\nonumber
&+& \int_{-\infty}^t d\,t' G_c(t-t')\sigma_+(t')\sigma_z(t) + n_+(t)\\ 
\label{es3} 
\dot{\sigma}_z(t) &=& i\Omega(\sigma_-(t)e^{i\Delta t} -\sigma_+(t)e^{-i\Delta t} + n_z(t)\\\nonumber
&&\hspace{-1.5cm} -2\int_{-\infty}^tdt' [G(t-t')\sigma_+(t)\sigma_-(t') + G_c(t-t')\sigma_+(t')\sigma_-(t)]
\end{eqnarray} 
Here $\Delta\equiv\omega_L-\omega_a$ and $\Delta_k\equiv\omega_a-\omega_k$.
The two functions $G(\tau)$ and $G_c(\tau)$ are the memory functions of the system caused by the atom-reservoir interaction and they are defined as $G(\tau) \equiv \sum_k |g_k|^2 e^{i\Delta_k\tau}\Theta(\tau)$, and $G_c(\tau) \equiv \sum_k |g_k|^2 e^{-i\Delta_k\tau}\Theta(\tau)$.
Here $\Theta(\tau)$ is the Heaviside step function.
The three noise operators $n_-(t)$, $n_+(t)$, and $n_z(t)$ originate from the original photon filed operator before interaction and are expressed as follows: 
\begin{eqnarray} 
&&n_-(t) = i\sum_k g_k e^{i\Delta_k t}\sigma_z(t) a_k({-\infty})\\
&&n_+(t) = - i\sum_k g_k^\ast e^{-i\Delta_k t}a_k^\dag({-\infty}) \sigma_z(t)\\
&&n_z(t) = -2 i\sum_k g_k e^{i\Delta_k t}\sigma_+(t) a_k({-\infty})\\\nonumber
&&\hspace{1cm} + 2 i\sum_k g_k^\ast e^{-i\Delta_k t}a_k^\dag({-\infty}) \sigma_-(t) 
\end{eqnarray} 
Supposing that the reservoir is in thermal equilibrium, it can be easily shown that at zero temperature the three noise operators are zero mean with only four non-zero correlation functions, $\langle n_-(t)n_+(t')\rangle_R$, $\langle n_-(t)n_z(t')\rangle_R$, $\langle n_z(t)n_+(t')\rangle_R$, and $\langle n_z(t)n_z(t')\rangle_R$.


To actually evaluate the memory functions as well as the correlation functions of the noise operators, one needs to know the spectral distribution of the photonic density of states.
Although in general the DOS of photonic bandgap crystals is very complicated and also varies with the geometrical structure and the dielectric constants of the material, it is still possible to approximately model the DOS near the bandedge with a simple formula.
According to the results from the full vectorial numerical calculation, the DOS near the bandgap for three-dimensional photonic crystals increases from zero and behaves more like the anisotropic model proposed in the literature \cite{SZhu00}.
To be more specific, for a three dimensional photonic bandgap crystal, if the wavevector that corresponds to the bandedge is ${\textbf k}_0^i$, then the dispersion relation in the anisotropic model is described by the following form:
$\omega_k=\omega_c+A|{\textbf k}-{\textbf k}_0^i|^2$, where $A$ is a model dependent constant and $\omega_c$ is the bandedge frequency.
Based on this dispersion relation, the corresponding DOS is given by: $ D(\omega) = \frac{1}{A^{3/2}}\sqrt{\omega-\omega_c}\Theta(\omega-\omega_c)$.
The memory functions under this anisotropic model also can be derived as: 
\begin{eqnarray} 
\label{Gpbg1} 
\tilde{G}(\omega) &=& \beta^{3/2}\frac{-i}{\sqrt{\omega_c}+\sqrt{\omega_c-\omega_a-\omega}}\\ 
\label{Gpbg2} 
\tilde{G_c}(\omega) &=& \beta^{3/2}\frac{i}{\sqrt{\omega_c}+\sqrt{\omega_c-\omega_a+\omega}} 
\end{eqnarray} 
where $\beta^{3/2}=\frac{\omega_a^2 d^2}{6\hbar\epsilon_0\pi A^{3/2}}\eta$, and we have used the space average coupling strength $\eta\equiv\frac{3}{8\pi}\int d\Omega|\hat{\textbf d}\cdot{\textbf E}|^2$ in the derivation.

\begin{figure} 
\begin{center}
\includegraphics[width=3.0in]{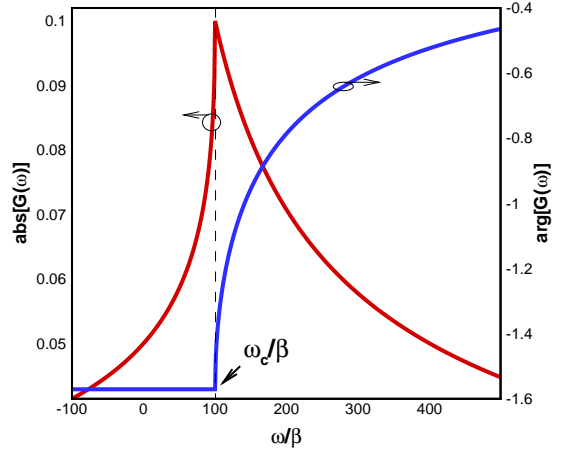} 
\caption{Amplitude and phase spectra of the memory function $G(\omega)$ with bandedge frequency $\omega_c = 100 \beta$. The memory function is non-uniform around the bandedge and becomes pure imaginary inside the bandgap.} 
\label{figgw}
\end{center}
\end{figure}

From Fig. \ref{figgw} the spectrum of the memory function $G(\omega)$ is non-uniform and asymmetric as we expect.
When the frequency is below the bandedge frequency $\omega_c$, the memory function becomes pure imaginary, indicating the inhibition of the spontaneous emission inside the bandgap.
The spectrum for another memory function $G_c(\omega)$ is also similar.
It can be easily checked that the full-width-half-maximum (FWHM) bandwidth of the memory functions in Eq. (\ref{Gpbg1}) and Eq. (\ref{Gpbg2}) are $4 \omega_c$. 
For the bandgap in optical domain, the order of $\omega_c$ is about $10^{14-15}$ Hz, and the typical lifetime of the atom is from $10^{-3}$ sec to $10^{-9}$ sec, which is much longer than the response time of the memory functions.

The generalized Bloch equations are a set of nonlinear operator equations and cannot be solved easily.
To overcome this difficulty, we introduce the following Liouville operator expansion: 
$\sigma_{ij}(t)=e^{-i{\cal L}(t-t')}\sigma_{ij}(t')=\sum_{n=0}^\infty \frac{[-i(t-t')]^n}{n!}{\cal L}^n\sigma_{ij}(t') $.
Here the Liouville super operator ${\cal L}$ is defined as 
${\cal L}^n \sigma_{ij}(t')=\frac{1}{\hbar^n}[\cdots[\sigma_{ij}(t'),H],H],\cdots,H]$.
If the atom we consider is with a longer lifetime and under weak pumping (small Rabi frequency) such as the usual case in optical frequency, the time scale of the atomic evolution will be always longer than the time scale of the memory functions in this case.
Therefore under such assumptions it should be legitimate to simply apply the zero-th order perturbation terms (zeroth-order Born approximation).
This is equivalent to use the equal time operator products to replace the two-time operator products.
With these approximations and by using the identities of Pauli matrices, the generalized optical Bloch equations can be reduced into the following form: 
\begin{eqnarray} 
\dot{\sigma}_-(t) &=& i\frac{\Omega}{2}\sigma_z(t)e^{-i\Delta t}\\\nonumber
&-&\int_{-\infty}^t d t' G(t-t')\sigma_-(t') + n_-(t)\\ 
\dot{\sigma}_+(t) &=& -i\frac{\Omega}{2}\sigma_z(t)e^{i\Delta t}\\\nonumber
&-&\int_{-\infty}^t d t' G_c(t-t')\sigma_+(t')+ n_+(t)\\ 
\dot{\sigma}_z(t) &=& i\Omega (\sigma_-(t)e^{i\Delta t} - \sigma_+(t)e^{-i\Delta t})\\\nonumber
&&\hspace{-1.0cm}- \int_{-\infty}^t d t' [G(t-t')+G_c(t-t')](1+\sigma_z(t')) + n_z(t) 
\end{eqnarray} 
This approximation should be valid as long as the time scale of the memory function is still much shorter than the time scale of the atomic response (i.e., the inverse of the Rabi frequency and the decay rate).
However, in contrast to the Markovian approximation, we do not approximate the memory function by a $\delta$-function, but instead still keep its finite response characteristics in order to include the effects due to the non-uniform density 
of states near the bandedge.
Although such a simple approximation includes only one portion of the non-Markovian nature of the problem, it should still be quite valid for the fluorescence spectrum calculation considered in the present work since here the memory function time scale is typically much shorter than the atomic response time scale.
We have also checked the validity of the zeroth-order approximation used here by carrying out the derivation in which the first order expansion term is also included (first-order Born approximation) \cite{MLewenstein87}.

Theoretically the fluorescence spectrum can be calculated by taking the Fourier transform of the first order correlation function of the atomic dipole moment operator.
When expressed in terms of the atomic correlation functions in the frequency domain, the expression for the fluorescence spectrum is given by: 
$ S(\omega) \propto \langle\tilde{\sigma}_+(\omega)\tilde{\sigma}_-(-\omega)\rangle_R $. 

\begin{figure}
\begin{center} 
\includegraphics[width=3.0in]{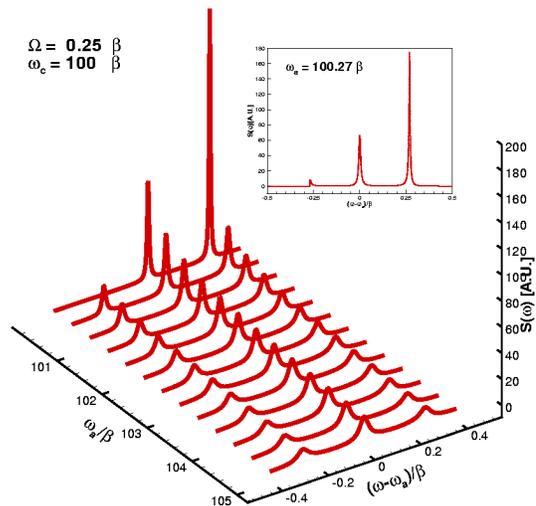} 
\caption{Resonance fluorescence spectrum near the bandedge at constant Rabi frequency.}
\label{swpbg1}
\end{center} 
\end{figure} 

Based on the above formula, in Fig. \ref{swpbg1} we plot the resonance fluorescence spectra at a constant Rabi frequency when the atomic transition frequency $\omega_a$ is near the bandedge, $\omega_c$.
All the frequency parameters used in the calculation are expressed in terms of the normalized frequency unit $\beta$ defined above and are labeled in the figure.
It can be noted that the separation of each adjacent peaks is also determined by the Rabi frequency as the case in free space.
When the atomic transition frequency is far away from the bandedge ($\omega_a\gg\omega_c$), the normal resonance fluorescence spectrum of Mollow's triplets is obtained just as expected.
However, as the atomic transition frequency moving toward the bandedge, the profiles of incoherent scattering processes become sharper and sharper as there are fewer and fewer DOS available.
The profile in the lower frequency is not only suppressed but also asymmetrical due to the existence of the bandgap and its residual profile exhibits a sharp edge as shown in the inset of the figure.
It should also be noted that the peak in the higher frequency is enhanced a lot as can be clearly seen in the figure.
Eventually the peak in the lower frequency will be totally suppressed when the atomic transition frequency is moving more toward the bandedge.
At this time the resonance fluorescence spectrum now only has two peaks.
More detailed results about the properties of resonance fluorescence intensity spectra will be reported elsewhere.  

\begin{figure}
\begin{center} 
\includegraphics[width=3.0in]{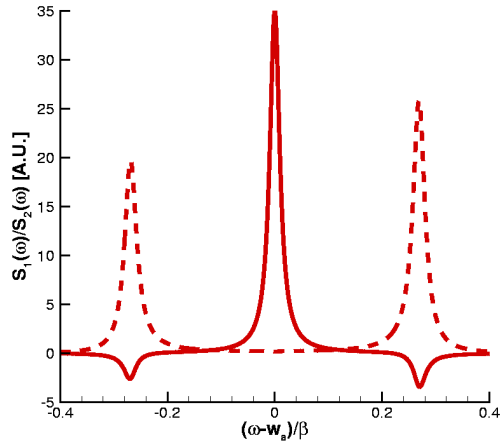} 
\caption{ Fluorescence quadrature spectra; Solid line: in-phase, $S_1(\omega)$, Dashed line: out-of-phase, $S_2(\omega)$ with $\omega_c = 100 \beta$ and $\Omega = 0.25 \beta$.}
\label{swpbg2}
\end{center} 
\end{figure} 

To observe squeezing, we need to calculate the fluorescence spectra for the quadrature field component. Theoretically the quadrature field operator in the $\theta$ phase angle is defined as $\hat{E}_\theta (t) = e^{i\theta} \hat{E}^{(+)}(t) + e^{-i\theta}\hat{E}^{(-)} (t)$.
The two cases corresponding to $\theta = 0$ and $\theta = \pi/2$ represent the in-phase and out-of-phase components of the electric field respectively.
The spectra for the quadrature field can be obtained by calculating the following normally order variance \cite{MCollett84, DWalls84}:
\begin{eqnarray}
&&S_\theta(\omega) \equiv <\tilde{E}_\theta(\omega),\tilde{E}_\theta(-\omega)> \\\nonumber
&&\propto \frac{1}{4}[<\tilde{\sigma}_-(\omega)\tilde{\sigma}_-(-\omega)>e^{-2 i \theta}+<\tilde{\sigma}_+(\omega)\tilde{\sigma}_-(-\omega)>\\\nonumber
&&+<\tilde{\sigma}_+(-\omega)\tilde{\sigma}_-(\omega)>+<\tilde{\sigma}_+(-\omega)\tilde{\sigma}_+(\omega)> e^{2 i \theta}]
\end{eqnarray}

\begin{figure}
\begin{center} 
\includegraphics[width=3.0in]{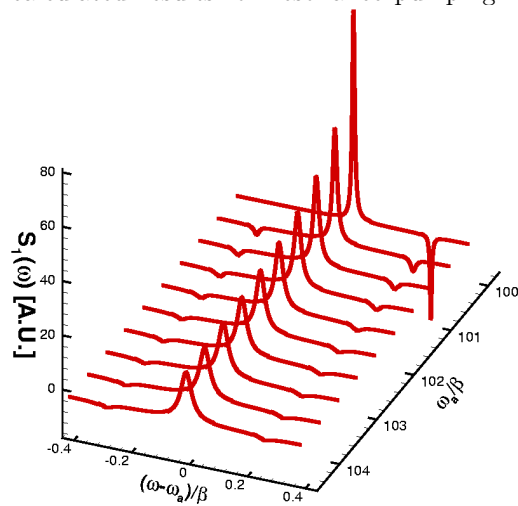} 
\caption{In-phase quadrature spectra near the bandedge with $\omega_c = 100 \beta$ and $\Omega = 0.25 \beta$.}
\label{swpbg3}
\end{center} 
\end{figure}

In the free space case, the in-phase quadrature $S_1(\omega)$ produces the central peak of the Mollow's triplet while the out-of-phase quadrature $S_2(\omega)$ produces the two side-peaks.
The situations are different for the case of photonic bandgap crystals.
When the transition frequency is near the bandedge of the photonic crystals, the bandgap effects modify the fluorescence intensity spectrum and cause asymmetric spectral profiles as we have seen in the Fig. (\ref{swpbg1}).
As shown in Fig. (\ref{swpbg2}), for the case of photonic bandgap crystals, the in-phase quadrature not only contributes to the central component but also the two sidebands. 
The out-of-phase quadrature still only contributes to the two sidebands as the case in free space.
Moreover, we find both sidebands of the in-phase quadrature now exhibit squeezing even when $\Omega^2>\frac{1}{4}\Gamma^2$.
The higher frequency peak exhibits larger squeezing as well as larger fluorescence intensity, Fig. (\ref{swpbg3}).

In conclusion, by introducing the Liouville operator expansion, we have successfully overcome some of the difficulties associated with the non-Markovian nature of the problem caused by the non-uniform distribution of the photon states in the photonic bandgap crystal. 
The calculated results for resonance pumping have indicated that the resonance fluorescence spectra near a photonic bandgap can exhibit interesting behaviors including the suppression and enhancement of the Mollow's triplet peaks, and the squeezing phenomena in the in-phase quadrature spectra.
It shall be very interesting to see if one can actually verify these predictions experimentally.

\end{document}